\documentclass[12pt,a4paper]{article}
\usepackage{amssymb}
\usepackage{amsfonts}
\usepackage{amssymb}
\usepackage{amsmath}
\usepackage{latexsym}
\usepackage{epsf,epsfig,rotating}
\usepackage{url}
\usepackage{multicol}
\usepackage{float}

\begin{document}

\title{Particle-like Q-balls}

\author{
 E.\,Ya.\;Nugaev$^a$\thanks{{\bf e-mail}:
emin@ms2.inr.ac.ru}, M.\,N.\;Smolyakov$^b$\thanks{{\bf e-mail}:
smolyakov@theory.sinp.msu.ru}
\\
$^a${\small{\em
Institute for Nuclear Research of the Russian Academy
of Sciences,}}\\
{\small{\em 60th October Anniversary prospect 7a, 117312, Moscow,
Russia
}}\\
$^b${\small{\em Skobeltsyn Institute of Nuclear Physics, Lomonosov
Moscow State University,
}}\\
{\small{\em  119991, Moscow, Russia}}}

\date{}
\maketitle

\begin{abstract}
Usually the charge and the energy of stable Q-balls vary in a wide
range or are even unbounded. In the present paper we study an
interesting possibility that this range is parametrically small.
In this case the spectra of stable Q-balls look similar to the
one of free particles.
\end{abstract}

Among the variety of non-topological solitons (see
\cite{Makhankov:1978rg,LeePang} for review) Q-balls
\cite{Rosen,Coleman:1985ki} and their properties were thoroughly
examined, in particular, due to the interest encouraged by
cosmology (see, for example, \cite{Gorbunov:2011zz}). The main
soliton characteristics, the energy $E$ and the charge $Q$, are
functions of the parameter $\omega$ (the standard Q-ball solution
in a scalar field theory with global $U(1)$ invariance has the
form $\phi(t,\vec{x})=f(|\vec{x}|) e^{i\omega t}$), which results
in the possibility of different forms of $E(Q)$ dependence for
different scalar field potentials.

Of course, the most interesting Q-ball solutions are stable
solutions. In general, there are three types of the Q-ball
stability:
\begin{enumerate}
\item The first type is the quantum mechanical
stability, i.e., the stability with respect to decay into free
particles. If $E(Q)<MQ$ for a Q-ball of charge $Q$, where $M$ is
the mass of a free particle in the theory under
consideration (without loss of generality, from here on we suppose
that $\omega\ge 0$ and $Q\ge 0$), then such a Q-ball is quantum
mechanically stable.\footnote{In the case of special interactions
with fermions this kinematic consideration should be revised, see
\cite{Cohen:1986ct}.}
\item
The second type is the stability against fission. Q-balls are
stable against decay into Q-balls with smaller charges if
$d^{2}E/dQ^{2}<0$ (a simple justification of this fact in the
general case can be found in \cite{Gulamov:2013ema}).
\item
The third type is the classical stability, i.e., the stability with
respect to small perturbations of the scalar field. The stability
criterion proposed in \cite{LeePang,Friedberg:1976me} implies that
a Q-ball is classically stable if $\frac{dQ}{d\omega}<0$.
\end{enumerate}
Below we will consider only those Q-ball solutions which satisfy
all the three stability criteria, presented above. We will call
them ``absolutely stable" Q-balls. It should be noted that since
the equality $\frac{dE}{dQ}=\omega$ always holds for Q-balls, the
latter leads to
\begin{equation}
\frac{d^2E}{dQ^2}=\frac{d\omega}{dQ}.
\end{equation}
Thus, the criterion of stability against fission and the criterion
of classical stability coincide, i.e., classically stable Q-balls
are stable against fission.

Note that our definition of the absolute stability (at least in the
absence of fermions) differs from the one of papers
\cite{Tsumagari:2008bv,Tsumagari:2009zp}, where the stability with
respect to decay into free particles is supposed to be the
strongest criterion, which Q-balls should satisfy, and such
Q-balls are called absolutely stable in these papers. Our
definition is different because, as we will see below, the stability
with respect to decay into free particles does not imply the classical
stability in the general case.

As it was noted above, the $E(Q)$ dependencies may have rather
different forms in models with different potentials. As the first
example one can recall the model presented in the well-known paper
\cite{Friedberg:1976me}. The $E(Q)$ dependence in this model
consists of two branches, one of which (the lower one) is
classically stable. Moreover, there exists $Q_{S}$ such that for
$Q>Q_{S}$ the inequality $E(Q)<MQ$ holds for the lower branch (see
Fig.~3(a) in \cite{Friedberg:1976me}). Thus, Q-balls with
$Q>Q_{S}$ from the lower branch of the $E(Q)$ dependence are
absolutely stable. An analogous form of the $E(Q)$ dependence is
inherent to other models, see, for example,
\cite{Gulamov:2013ema,Alford:1987vs,Theodorakis:2000bz,Tamaki:2012yk}.

Another type of Q-balls is the one with only one branch. As an
example one may consider the model with $|\phi|^4$ potential
studied in \cite{Anderson:1970et}. The $E(Q)$ dependence in this
model consists of only one branch with $d^{2}E/dQ^{2}>0$, and all
Q-balls in such a model are even classically unstable (this was
also shown explicitly in \cite{Anderson:1970et}).

An interesting model with a logarithmic unbounded\footnote{Surely,
one can add positive terms to the potential for very large values
of the field modulus without altering the physics at the scale of
stable Q-balls.} scalar field potential was proposed in
\cite{Rosen1} and thoroughly examined in \cite{MarcVent}. The
$E(Q)$ dependence in this model also consists of two branches, one
of which is classically stable (again it is the lower branch). The
charge of the Q-balls from the stable branch varies from $0$ to
$Q_{max}<\infty$. An analogous $E(Q)$ behavior has the model with
a simple polynomial potential discussed in \cite{Tamaki:2012yk}.

In all the examples presented above the spectra of stable Q-balls
(if they exist) either have no upper limit, or have an upper
limit, but start from zero. In any case, the charge and the energy
of such Q-balls vary in a wide range or are even unbounded. There
arises a question: is it possible to make this range
parametrically small?

To answer this question, we recall that in some models there is
another form of the $E(Q)$ dependence. It consists of three
branches, one of which, -- the ``lowest" branch, contains Q-balls
which are classically stable. An important feature of this branch
is that there exist both a lower bound on the charge $Q_{min}$ and
an upper bound $Q_{max}$ such that Q-balls with
$Q_{min}<Q<Q_{max}$ are classically stable. Such an $E(Q)$
dependence arises in the models with piecewise parabolic
potentials examined in \cite{Gulamov:2013ema,Theodorakis:2000bz}
(these scalar field potentials were originally proposed in \cite{Rosen}),
and in the model with a polynomial potential discussed in
\cite{Tamaki:2012yk}. Below we will focus on examination of such
an $E(Q)$ dependence with three branches.

In order to find out whether it is possible that the range of
charges, where the absolutely stable Q-balls exist, can be made
small, it is better to have an analytically solvable model. The
models discussed in \cite{Gulamov:2013ema,Theodorakis:2000bz} are
analytically solvable (the model of \cite{Gulamov:2013ema}
provides a very simple analytic Q-ball solution, which is very
useful for examining perturbations above the Q-ball solution
explicitly), but the scalar field potentials utilized in these
models contain breaks, which is rather unphysical and demands an
additional regularization of the potentials. Below we will propose
a model with a continuous and differentiable potential, admitting
the existence of a simple analytic Q-ball solution and providing
the $E(Q)$ dependence with three branches, one of which
corresponds to classically stable Q-balls. We will calculate $Q_S$
and $Q_{max}$ in this model and answer the question posed above.

\begin{figure}[ht]
\includegraphics[width=6.5in]{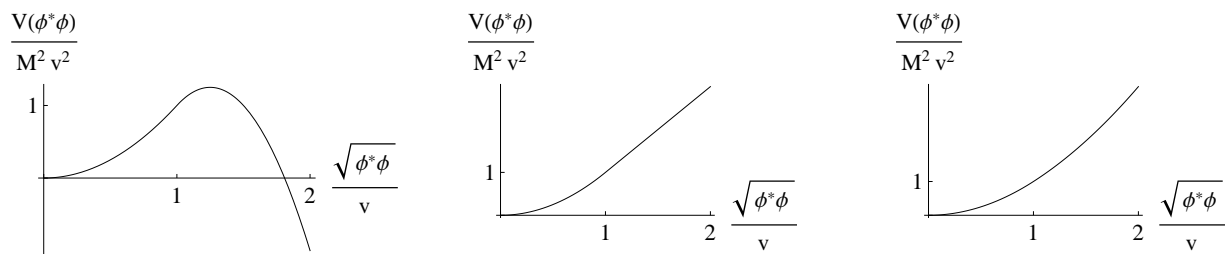}
\caption{The forms of the scalar field potential described by
Eq.~(\ref{potential}): $m^2<0$, $|m|/M=2$ (left plot); $m=0$
(middle plot); $m^2>0$, $m/M=0.9$ (right plot).} \label{V}
\end{figure}

We consider the globally $U(1)$ invariant scalar field theory with
a piecewise potential of the form
\begin{eqnarray}\label{potential}
V(\phi^{*}\phi)&=&M^2\phi^{*}\phi\,\theta(v^{2}-\phi^{*}\phi)\\
\nonumber&+&\left(m^2\phi^{*}\phi+2v(M^2-m^2)\sqrt{\phi^{*}\phi}-v^2(M^2-m^2)\right)\theta(\phi^{*}\phi-v^{2}),
\end{eqnarray}
where $M^2>0$, $\theta$ is the Heaviside step function with the
convention $\theta(0)=\frac{1}{2}$. The form of this scalar field
potential for different values of the dimensionless parameter
$\frac{m}{M}$ is presented in Fig.~\ref{V}.
We will be looking for a solution to the corresponding equation of
motion of the standard form $\phi=f(r,\omega)e^{i\omega t},$ where
$r=|\vec x|$. Without loss of generality, we suppose that
$f(r,\omega)>0$. The monotonic solution for $f$ such that
$\frac{df}{dr}\bigl|_{r=0}=0$ and $f|_{r\to\infty}=0$ can be
easily found and has the form
\begin{eqnarray}\label{backgr1}
f(r,\omega)=v\frac{(M^2-m^2)}{(\omega^2-m^2)}-v\frac{(M^2-\omega^2)}{(\omega^2-m^2)}\frac{R}{r}\frac{\sin(\sqrt{\omega^2-m^2}r)}{\sin(\sqrt{\omega^2-m^2}R)},\qquad
r<R,\\ \label{backgr2}
f(r,\omega)=v\frac{R}{r}\frac{e^{-\sqrt{M^2-\omega^2}r}}{e^{-\sqrt{M^2-\omega^2}R}},\qquad
r\ge R,
\end{eqnarray}
where the matching radius $R$ is such that $f(R,\omega)=v$. For $r<R$ we have
$f(r,\omega)>v$, whereas for $r>R$ we have $f(r,\omega)<v$. It is evident that
if $m^2>0$, then $M>\omega>m$; if $m=0$, then $M>\omega>0$;
otherwise $M>\omega\ge 0$.

The continuity of $f(r,\omega)$ and of its first derivative leads to the
following equation for $R=R(\omega)$:
\begin{eqnarray}\label{defR}
\left(\frac{M^2-m^2}{\omega^2-m^2}+\sqrt{M^2-\omega^2}R\right)\tan(\sqrt{\omega^2-m^2}R)=\frac{M^2-\omega^2}{\sqrt{\omega^2-m^2}}R.
\end{eqnarray}
This equation can be easily solved numerically for a given
$\omega$. Note that equation (\ref{defR}) is valid only for Q-ball
solutions without nodes. For such a solution and for a given
$\omega$ one should take the first (smallest) root of (\ref{defR})
satisfying the condition
$R(\omega)>\frac{\pi}{\sqrt{\omega^2-m^2}}$.

The Q-ball charge and energy can also be easily calculated and
have the form
\begin{eqnarray}\label{chargeorig}
Q=2\omega\int\limits_{0}^{\infty}f^2d^3x=\\ \nonumber 4\pi \omega
v^2\left[\frac{R^3}{(\omega^2-m^2)^2}\left(\frac{2}{3}(M^2-m^2)^2+(M^2-\omega^2)^2+(\omega^2-m^2)(M^2-\omega^2)\right)\right.\\
\nonumber\left.+\frac{R^2\sqrt{M^2-\omega^2}}{(\omega^2-m^2)^2}\left(5M^2-6m^2+\omega^2)\right)+5\frac{R(M^2-m^2)}{(\omega^2-m^2)^2}+
\frac{R^2}{\sqrt{M^2-\omega^2}}\right],
\end{eqnarray}
\begin{eqnarray}
E=\omega Q+4\pi
v^2\frac{M^2-m^2}{\omega^2-m^2}\left[\frac{R^3}{3}(M^2-\omega^2)+R^2\sqrt{M^2-\omega^2}+R\right],\label{energyorig}
\end{eqnarray}
where we have used Eq. (\ref{defR}) in the derivation.

Now let us examine the $E(Q)$ dependence for different values of
the model parameters, i.e., for $m^2>0$, $m=0$ and $m^2<0$. It is not difficult to show that the charge (\ref{chargeorig}) and the energy (\ref{energyorig}) can be represented as
\begin{eqnarray}
Q=\frac{4\pi v^{2}}{M^2}\tilde Q,\qquad
E=\frac{4\pi v^{2}}{M}\tilde E,
\end{eqnarray}
where $\tilde Q$ and $\tilde E$ depend only on $\frac{\omega}{M}$ and $\frac{m}{M}$ and do not depend on $v$. So, below we will not specify the values of $v$ and $M$ while examining the main properties of Q-balls in our model: the $E(Q)$ dependencies can be examined by considering the dimensionless quantities $\tilde Q$ and $\tilde E$ for different choices of $\frac{m}{M}$. Such a simplification is possible only because of the simple form of the scalar field potential, which appears to be very useful for calculations.

\begin{figure}[H]
\begin{center}
\includegraphics[width=14cm]{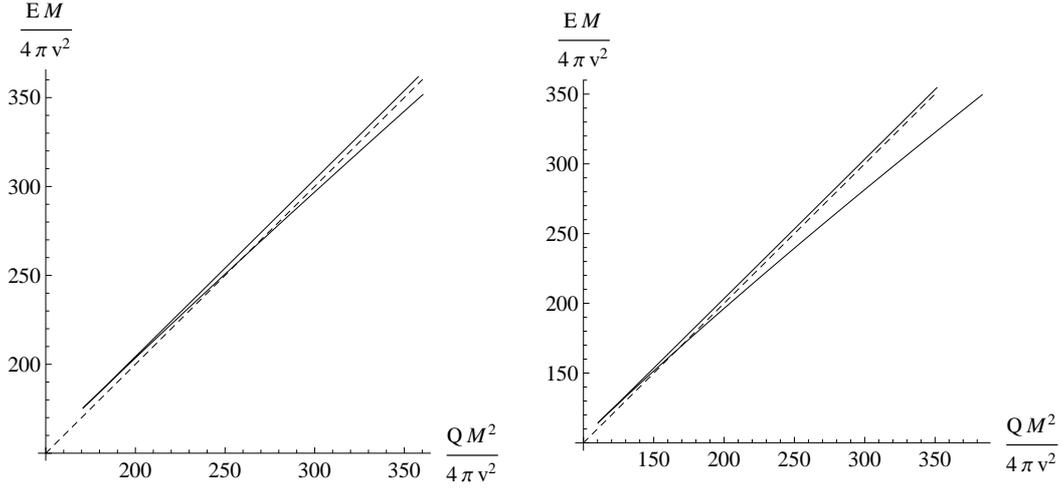}\caption{E(Q) for $m^2>0$, $\frac{m}{M}=0.5$ (left plot) and for $m=0$ (right plot).
The dashed line corresponds to free scalar particles of mass
$M$.}\label{EQf1}
\end{center}
\end{figure}

\begin{figure}[H]
\begin{center}
\includegraphics[width=14cm]{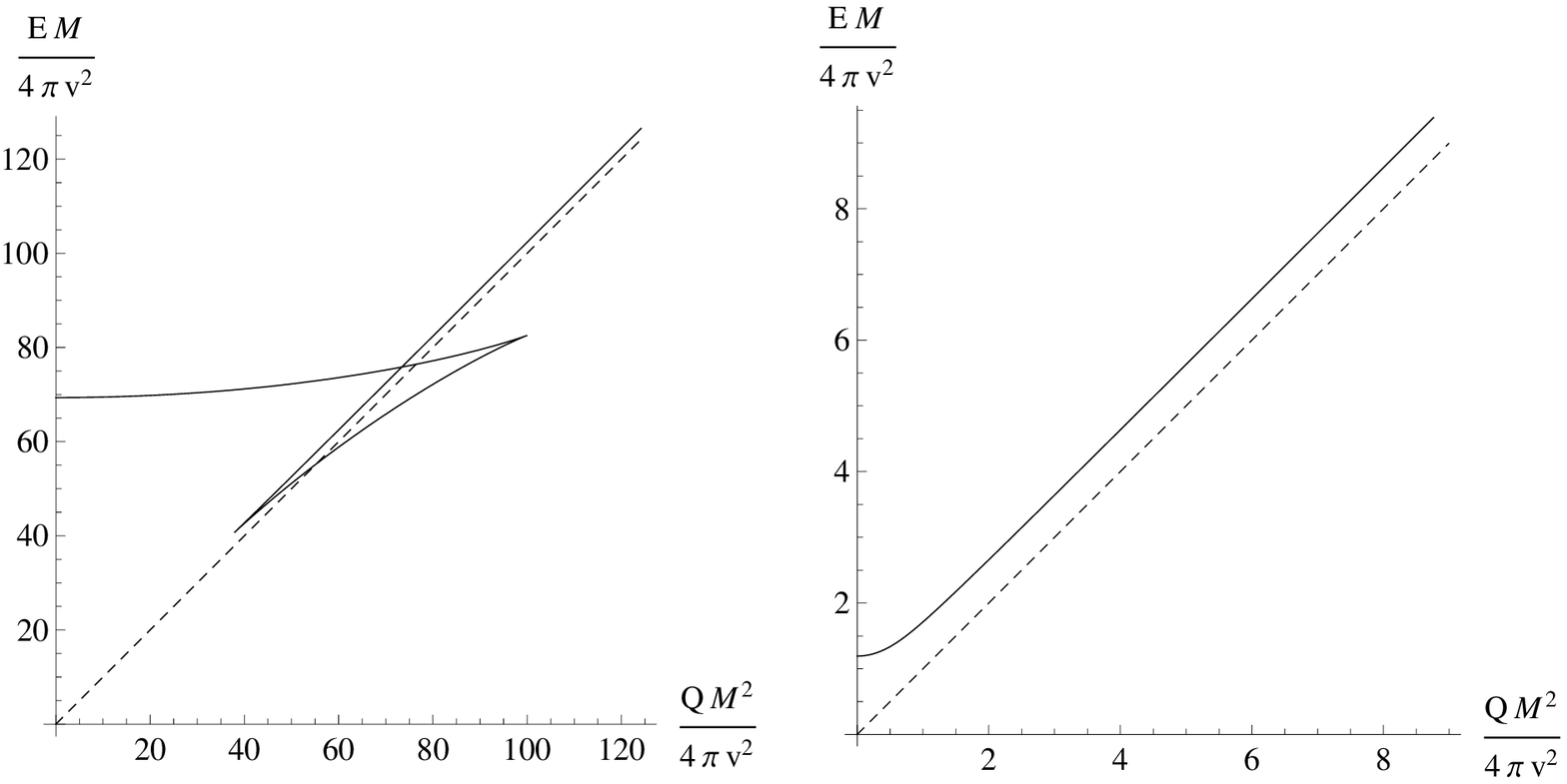}\caption{E(Q) for $m^2<0$. $\frac{|m|}{M}=1$ (left plot) and $\frac{|m|}{M}=5$ (right plot).
The dashed line corresponds to free scalar particles of mass
$M$.}\label{EQf2}
\end{center}
\end{figure}

The corresponding plots are presented in Figs.~\ref{EQf1}~and~\ref{EQf2}.
We see that the $E(Q)$ diagrams
for the cases $m^2>0$ and $m=0$ resemble those in the models
discussed in
\cite{Friedberg:1976me,Alford:1987vs,Theodorakis:2000bz,Tamaki:2012yk}.
All four cases, presented in Figs.~\ref{EQf1}~and~\ref{EQf2}, also
exist in the model discussed in \cite{Gulamov:2013ema}.
It should be noted that, though $\omega$ is bounded from above, $\omega<M$, in the limit $\omega\to M$ the charge and the energy tend to infinity in all four cases, presented in Figs.~\ref{EQf1}~and~\ref{EQf2}. This happens because the factor $\sqrt{M^{2}-\omega^{2}}$ in the exponent of (\ref{backgr2}) tends to zero for $\omega\to M$, whereas $R(\omega)|_{\omega\to M}\to\frac{\pi}{\sqrt{M^2-m^2}}$; so the scalar field falls off not exponentially, but as $\frac{1}{r}$ in this limit. The latter leads to infinite charge and energy of the Q-ball for $\omega\to M$. Due to the large size of the Q-ball core, such Q-balls were called ``Q-clouds'' in \cite{Alford:1987vs}.

As it was noted above, we will be interested in the last case $m^2<0$. As it can be seen from Fig.~\ref{EQf2}, there are two
phases: the first phase contains three branches on the $E(Q)$
diagram, whereas the other phase contains only one branch (the
latter case is similar to the one of the model with $|\phi|^4$
potential studied in \cite{Anderson:1970et}). The transition
between the phases occurs at $\frac{|m|}{M}\approx 1.775$. One
also sees from the left plot in Fig.~\ref{EQf2} that the most part
of the lowest stable branch lies under the $E=MQ$ line
corresponding to free particles, which means that the range of
charges of absolutely stable Q-balls is rather large. Note that
the part of the upper classically unstable branch (which starts
from $Q=0$) on the left plot in Fig.~\ref{EQf2} also lies under
the $E=MQ$ line corresponding to free particles, which means that the
stability with respect to decay into free particles indeed does
not imply the classical stability in the general case.

We would like to note that the existence of a locally maximal charge in the phase with three branches (see Fig.~\ref{EQf2}) seems to be a consequence of the turnover of the scalar field potential. We think that this is a rather general property, which is inherent to other models of Q-balls. Although we can not prove it in a rigorous way, we do not know exceptions from this rule. Meanwhile, the opposite is not correct --- the existence of the turnover of the scalar field potential does not guarantee the existence of a locally maximal charge, which is confirmed by the existence of the phase without maximal charge for $\frac{|m|}{M}>1.775$ in our case and by the examples of other models (see, for example, \cite{Anderson:1970et}). We also stress that the maximal value of $f(r,\omega)$ (which is simply $f(0,\omega)$) of the Q-ball with locally maximal charge is not connected with the point of the maximum of the scalar field potential for $m^{2}<0$. Indeed, the scalar field potential is maximal at $f_{\textrm{Vmax}}=v\left(1+\frac{M^2}{|m|^2}\right)$; whereas the value of $f(0,\omega)$ decreases monotonically (this can be checked numerically) from $f(0,0)$ to $f(0,M)=2v$. So, if $\frac{|m|}{M}\ge 1$ (see, for example, left plot in Fig.~\ref{EQf2}), then $f(0,\omega)>2v\ge f_{\textrm{Vmax}}$: the maximum of the absolute value of the Q-ball scalar field is larger than the point of the maximum of the scalar field potential for any $0\le\omega<M$, i.e., for any Q-ball. An interesting observation in the opposite case $\frac{|m|}{M}<1$ is that Q-balls with $\omega\to M$ lie on the unstable branch, whereas $f(0,M)=2v<f_{\textrm{Vmax}}$ in this case. These examples demonstrate that there is no (at least obvious) connection between the maximal absolute value of the Q-ball scalar field, the point of the maximum of the scalar field potential and the Q-ball stability.

\begin{figure}[ht]
\includegraphics[width=16cm]{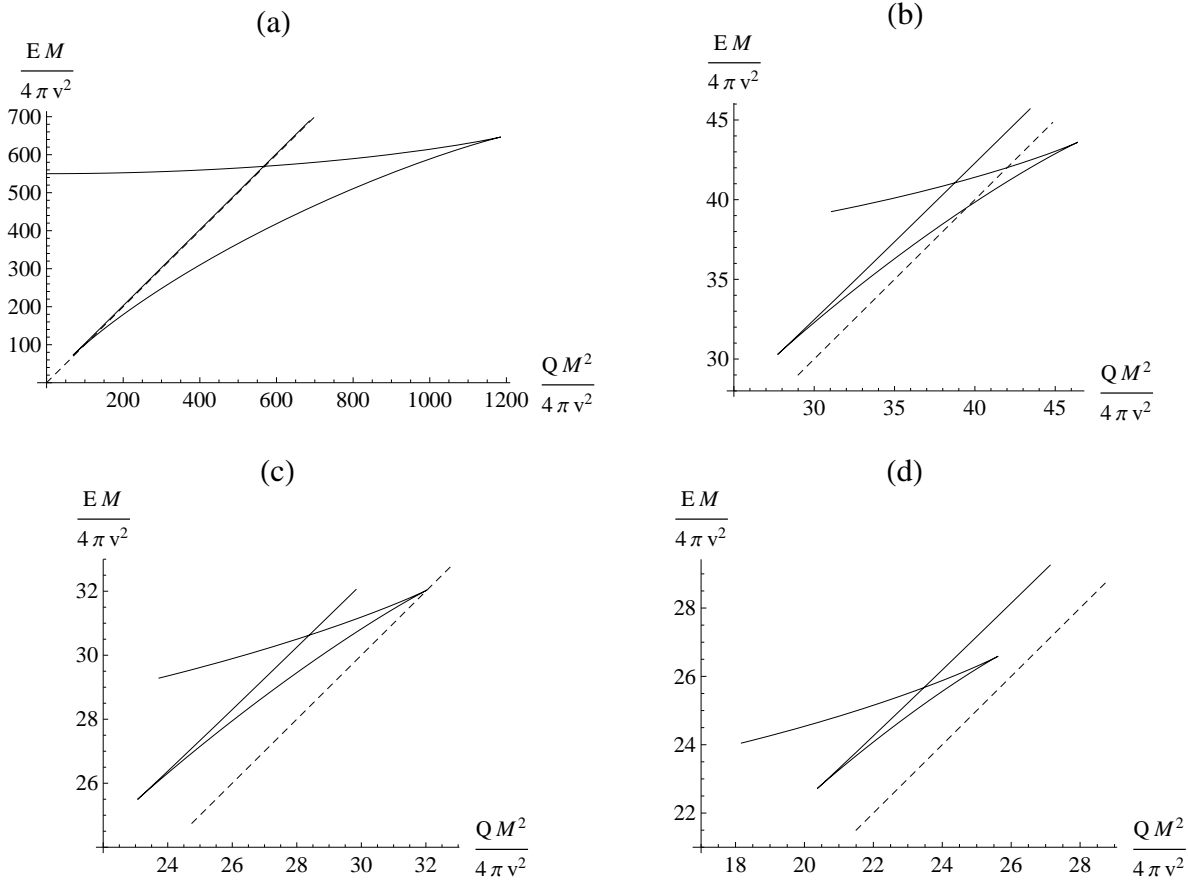}\caption{E(Q) for $m^2<0$. $\frac{|m|}{M}=0.6$ (a);
$\frac{|m|}{M}=1.2$ (b); $\frac{|m|}{M}=1.31886$ (c);
$\frac{|m|}{M}=1.4$ (d). The dashed line stands for free particles
of mass $M$.}\label{EQf3}
\end{figure}
Now let us check what happens when we change the parameter $\tilde
m=\frac{|m|}{M}$. The result is presented in Fig.~\ref{EQf3}. We
see that the larger $\tilde m$ is, the smaller the ``triangle" in
the corresponding $E(Q)$ diagram is. Moreover, the larger $\tilde
m$ is, the smaller part of this ``triangle" turns out to lie under
the $E=MQ$ line corresponding to free particles. For $\tilde
m=\tilde m_{x}\approx 1.31886$ the ``triangle" touches the free
particles line by the upper cusp, whereas for $\tilde m>\tilde
m_{x}$ all the classically stable Q-balls are quantum mechanically
unstable. For $\tilde m \gtrsim 1.775$ the ``triangle" disappears
and there is no classically stable branch in the $E(Q)$ dependence
at all.

\begin{figure}[ht]
\begin{center}
\includegraphics[width=16cm]{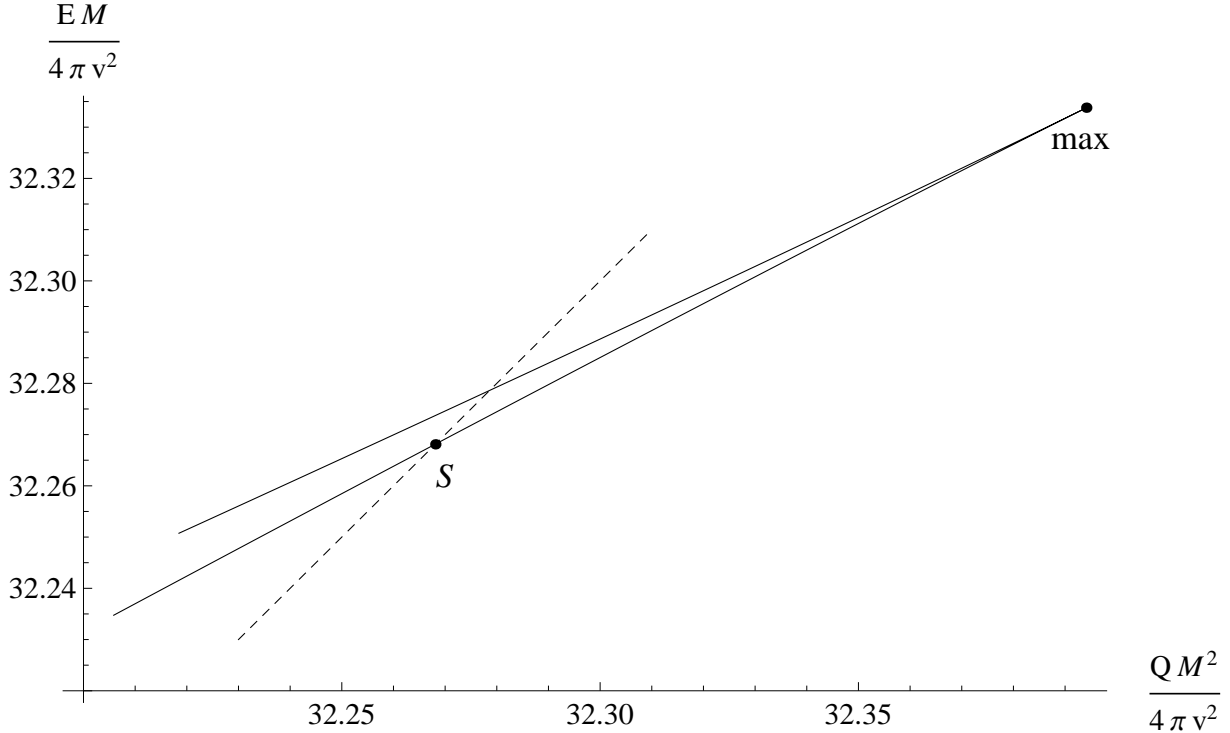}\caption{E(Q) for $m^2<0$. $\frac{|m|}{M}=1.315$.
The dashed line stands for free particles of mass
$M$.}\label{EQf4}
\end{center}
\end{figure}

The observations presented above indicate that there exist such
values of the parameters that absolutely stable Q-balls can exist
only in a very narrow range of charges (and, consequently,
energies). As an example, let us take $\tilde m=1.315$. The upper
right part of the ``triangle", including the cusp, is presented in
Fig.~\ref{EQf4}.
The lower branch in this figure is the stable one. The values of
$\tilde Q=\frac{Q M^2}{4\pi v^2}$ and $\tilde E=\frac{E M}{4\pi
v^2}$, corresponding to the dots on the plot, are the following:
$\tilde Q_{S}\approx 32.268$, $\tilde E_{S}\approx 32.268$,
$\tilde Q_{max}\approx 32.394$, $\tilde E_{max}\approx 32.334$.
The ranges of the charges and the energies, for which the absolutely stable
Q-balls can exist, are $$\Delta\tilde Q=\tilde Q_{max}-\tilde
Q_{S}\approx 0.126,\qquad \Delta\tilde E=\tilde E_{max}-\tilde
E_{S}\approx 0.066.$$ These ranges are much smaller than the absolute
values of the charges and the energies respectively.

The closer (from below) $\tilde m$ to $\tilde m_{x}\approx
1.31886$ is, the smaller $\Delta\tilde Q$ and $\Delta\tilde E$
are. For $\Delta\tilde Q\ll\tilde Q_{max}$ the $E(Q)$ dependence
of the absolutely stable Q-balls is similar to the one in the
limiting case $\tilde m\to \tilde m_{x}$:
$$E=MQ_{x},$$ where $\tilde Q_{x}=\tilde Q_{max}|_{\tilde m=\tilde
m_{x}}\approx 32.034$. But it looks exactly like the $E(Q)$
dependence of free particles at rest! The only difference is that
the charge of free particles $Q_{p}=1$, whereas for Q-balls we
have in the limiting case $Q_{x}\approx 32\frac{4\pi v^2}{M^2}$.
Of course, analogous anti-Q-balls (i.e., Q-balls with $\omega<0$
and $Q<0$) also exist and possess the same properties.

We think that the existence of such particle-like Q-balls should
be inherent not only to the model presented above, but to other
models providing an $E(Q)$ dependence with three branches (namely,
to models with scalar field potential admitting the existence of a
true vacuum at $\phi^{*}\phi>0$ or at least having a negative
slope after some nonzero value of the scalar field modulus), like
those in \cite{Gulamov:2013ema,Theodorakis:2000bz,Tamaki:2012yk}.
In such models there is a possibility to tune the charge of a
stable Q-ball to a nearly determined value, i.e., it is possible
to have absolutely stable Q-balls with the ranges of charges and
energies much smaller than the absolute values of charges and
energies themselves. The spectra of such Q-balls are very similar
to the spectrum of free particles of the theory and Q-balls behave
like clusters of free particles, which looks very intriguing.

\section*{Acknowledgements}
The authors are grateful to D. Levkov, M. Libanov and I. Volobuev
for discussions and to the unknown referee for useful comments. The work was supported by RFBR grant 14-02-31384. The work of E.Y.N.
was supported in part by grant NS-2835.2014.2 of the President of
Russian Federation and by RFBR grant 13-02-01127a. The work of
M.N.S. was supported in part by grant NS-3042.2014.2 of the
President of Russian Federation and by RFBR grant
12-02-93108-CNRSL-a.


\begin{thebibliography}{99}
\bibitem{Makhankov:1978rg}
  V.~G.~Makhankov,
  %``Dynamics of Classical Solitons In Nonintegrable Systems,''
  Phys.\ Rept.\  {\bf 35} (1978) 1.

\bibitem{LeePang}
T.~D.~Lee and Y.~Pang,
  %``Nontopological solitons,''
  Phys.\ Rept.\  {\bf 221} (1992) 251.

\bibitem{Rosen}
G. ~Rosen,
J.\ Math.\ Phys. {\bf 9} (1968) 996.

\bibitem{Coleman:1985ki}
  S.~R.~Coleman,
  %``Q Balls,''
  Nucl.\ Phys.\ B {\bf 262} (1985) 263
   [Erratum-ibid.\ B {\bf 269} (1986) 744].

\bibitem{Gorbunov:2011zz}
  D.~S.~Gorbunov and V.~A.~Rubakov,
``Introduction to the theory of the early universe: Hot big bang
theory", Hackensack, USA: World Scientific (2011), 473 p.

\bibitem{Cohen:1986ct}
  A.~G.~Cohen, S.~R.~Coleman, H.~Georgi and A.~Manohar,
  %``The Evaporation Of Q Balls,''
  Nucl.\ Phys.\ B {\bf 272} (1986) 301.

\bibitem{Gulamov:2013ema}
  I.~E.~Gulamov, E.~Y.~Nugaev and M.~N.~Smolyakov,
  %``Analytic Q-ball solutions and their stability in a piecewise parabolic potential,''
  Phys.\ Rev.\ D {\bf 87} (2013) 085043.

\bibitem{Friedberg:1976me}
  R.~Friedberg, T.~D.~Lee and A.~Sirlin,
  %``A Class of Scalar-Field Soliton Solutions in Three Space Dimensions,''
  Phys.\ Rev.\ D {\bf 13} (1976) 2739.

\bibitem{Tsumagari:2008bv}
  M.~I.~Tsumagari, E.~J.~Copeland and P.~M.~Saffin,
  %``Some stationary properties of a Q-ball in arbitrary space dimensions,''
  Phys.\ Rev.\ D {\bf 78} (2008) 065021.

\bibitem{Tsumagari:2009zp}
  M.~I.~Tsumagari,
  %``The Physics of Q-Balls,''
  arXiv:0910.3845 [hep-th].

\bibitem{Alford:1987vs}
  M.~G.~Alford,
  %``Q Clouds,''
  Nucl.\ Phys.\ B {\bf 298} (1988) 323.

\bibitem{Theodorakis:2000bz}
  S.~Theodorakis,
  %``Analytic Q ball solutions in a parabolic-type potential,''
  Phys.\ Rev.\ D {\bf 61} (2000) 047701.

\bibitem{Tamaki:2012yk}
  T.~Tamaki and N.~Sakai,
  %``Unified pictures of Q-balls and Q-tubes,''
  Phys.\ Rev.\ D {\bf 86} (2012) 105011.

\bibitem{Anderson:1970et}
D.~L.~T.~Anderson and G.~H.~Derrick,
%``Stability of time-dependent particlelike solutions in nonlinear field theories. 1.,''
 J.\ Math.\ Phys.\  {\bf 11} (1970) 1336.

\bibitem{Rosen1}
  G.~Rosen,
  %``Dilatation covariance and exact solutions in local relativistic field theories,''
  Phys.\ Rev.\  {\bf 183} (1969) 1186.

\bibitem{MarcVent}
G.~C.~Marques and I.~Ventura,
  %``Resonances Within Nonperturbative Methods in Field Theories,''
  Phys.\ Rev.\ D {\bf 14} (1976) 1056.

\end{thebibliography}
\end{document}